\documentclass[twocolumn,showpacs,preprintnumbers,amsmath,amssymb]{revtex4}

\usepackage{epsf}
\usepackage{graphicx}  
\usepackage{dcolumn}   
\usepackage{bm}        

\newcommand{\be}{\begin{equation}}
\newcommand{\en}{\end{equation}}
\newcommand{\bea}{\begin{eqnarray}}
\newcommand{\ena}{\end{eqnarray}}

\begin{document}

\preprint{BNU/0015-2006}

\title{Curvaton reheating mechanism in inflation on warped Dvali-Gabadadze-Porrati brane }

\author{Hongsheng Zhang  and
  Zong-Hong Zhu \footnote{E-mail address: zhuzh@bnu.edu.cn}}
\affiliation{
 Department of Astronomy, Beijing Normal University, Beijing 100875, China}
\date{ \today}

\begin{abstract}

 An impressed feature of inflation on warped Dvali-Gabadadze-Porrati (DGP) brane
 is that the inflationary phase  exits spontaneously for a scalar
 inflaton field with exponential potential, which presents a graceful exit mechanism
 for the inflation. But its reheating mechanism  leaves
 open.  We investigate the curvaton reheating  in inflation on warped DGP
  brane model. The reheating may occur in effctively 5 dimensional or 4 dimensional
  stage. We study the permitted parameter space of the curvaton
  field in detail. We demonstrate how the inflation
  model of the warped DGP brane is improved by the curvaton mechanism.

\end{abstract}

\pacs{ 98.80. Cq } \keywords{inflation, reheating}

\maketitle

\section{Introduction}

 The inflation is a work schema, rather than a model. There
 are several inflation models since Guth's seminal work \cite{guth}, for a
 review, see, for example \cite{review}. Though the inflationary
 scenario  successfully solved several problems in the standard big bang model,
 how to get an inflaton field from fundamental field theory keeps
 unsolved. In view of achievements and shortcomings of inflationary scenario,
 it is necessary to further our understanding of the
inflationary scenario from a theoretical perspective.

 The brane world scenario, in which the standard model particles are confined
 on the 3-brane while the gravitation can propagate in the whole
 space, is an important progress in high energy physics. Among various
 brane universe models, the one proposed in
 ~\cite{dgpmodel}, called DGP model, is very interesting.
 In the DGP model, gravity appears
 4-dimensional at short distances but is altered at distance large
 compared to some freely adjustable crossover scale $r_0$ through the
 slow evaporation of the graviton off our 4-dimensional brane world
 universe into an unseen, yet large, fifth dimension. The original
 DGP model was soon generalized to warped DPG model \cite{eff}. In some sense,
 it is a hybrid model sharing some features of Randall-Sundrum model \cite{rs} and
 some features of DGP model.

  The inflation on warped
  DGP brane has been
  discussed in \cite{pz} and \cite{cai}. It is found that there
  may exist three stages in the inflation period: At the ultra high
  energy limit, the spacetime is effectively 4-dimensional; at middle
  energy region, the spacetime is effectively 5-dimensional; and
  then the spacetime undergoes the second 4-dimensional stage before
  nucleosynthesis (note that in the ultra low energy limit, such as the
  present day, the model can enter another 5-dimensional stage
  again). In a concrete numerical example of warped DGP inflation
  model, the universe starts to inflate at the 4-dimensional stage and ends at
  5-dimensional stage \cite{cai}. In inflation scenarios, the exponential potential is an important
 example which can be solved exactly in the standard model. In
 addition, we know that such exponential potentials of scalar fields
 occur naturally in some fundamental theories such as string/M
 theories. It is worth noting that in this inflation model based on the warped DGP brane
world scenario, for an exponential potential, the inflationary phase
can exit naturally. In this inflationary scenario, when the energy
density decreases, the scalar field makes a transition into a
kinetic energy dominated regime, which means the slow roll
parameters becomes larger than 1, bringing inflation to an end.
Since the inflaton survives this process without decay, an
alternative reheating mechanism is required.

  It is suggested in \cite{gra} that reheating took place via
gravitational particle production. This particle production
mechanism roots in the different vacuums of the inflationary phase
and the kinetic phase. The different vacuums are related by
Bogoliubov
  transformations. The vacuum state in the inflationary phase is no
  longer vacuum state measured in the kinetic phase: There are
  particles generated. However this mechanism is very inefficiency.
  So there is a rather long kinetic energy
 dominated regime, which leads to short-wavelength gravitational waves
 to reach excessive amplitudes \cite{sss}. The gravitational particle
 production mechanism is also plagued by its prediction on perturbation
 spectrum: It gives an spectrum far from scale invariant, while the
 observation data prefer a nearly scale invariant spectrum.

 In this paper we shall propose a curvaton reheating mechanism for
 the warped DGP inflation model. The curvaton mechanism is firstly
 suggested as an alternative mechanism to generate the primordial
 scalar perturbation which is responsible for the structure formation.
  In the curvaton scenario the primordial density perturbation originates from the vacuum
 fluctuation of some ``curvaton'' field $\sigma$, different from the
  inflaton field \cite{curvaton}. The curvaton reheating mechanism was
  firstly suggested in context of the quintessential inflation model in \cite{fengli}.
  It is shown in \cite{dimo} that the eta-problem of quintessential inflation
   can be  also ameliorated under the curvaton hypothesis.
  And recently the curvaton reheating mechanism is applied to the different inflation
  models \cite{inf}.

  The organization of this paper is as follows. For a thorough discussion
  on the curvaton reheating process, in the next section
  we briefly review the warped DGP inflation, stressing on the different stages
  of different effective dimensions. In this section we also reanalyze the example discussed
  in \cite{cai}, pointing out the critical points of the dimension
  transition, which is very important for the studies on the permitted
  parameter space of the curvaton field. In section III we traverse
   the constraint on the curvaton field. We
  investigate all the possible 3 cases: 1. The curvaton oscillates and
  decays in 5-dimensional stage; 2. the curvaton oscillates in 5-dimensional stage
  but decays in 4-dimensional stage; and 3. the curvaton oscillates and
  decays in 4-dimensional stage. In the last section we present our
  conclusions and discussions.

 \section{A short review on the warped DGP inflation model}
 For further discussions on curvaton reheating after inflationary phase, we
 first give a brief review on the warped DGP inflation model. For a
 thorough discussion, see \cite{cai}.

 We start from the Friedmann equation on the warped DGP brane.
 Assuming a Friedmann-Robertson-Walker (FRW) metric on the brane, one derives the
 Friedmann equation,
 \be
 \label{fried}
 H^2+{k\over a^2}={1\over 3\mu^2}\Bigl[\,\rho+\rho_0\bigl(1 +
\epsilon{\cal A}(\rho, a)\bigr)\,\Bigr],
 \en
 where as usual $H$ is the Hubble parameter, $a$ denotes the scale factor,
 $\rho$ denotes the matter field density,
 $k$ is the constant curvature of the maximal symmetric
 space of the FRW metric, $\mu$ is a parameter with dimension of mass
 , which denotes the strength of the induced gravity term on the brane,
   and $\epsilon$ denotes either $+1$ or $-1$, represents
 the two branches of this model.  ${\cal A}$ is
defined by

\be
 {\cal A}=\left[{\cal A}_0^2+{2\eta\over
\rho_0}\left(\rho-\mu^2 {{\cal E}_0\over a^4} \right)\right]^{1\over
2},
 \en
where
 \be
 {\cal A}_0=\sqrt{1-2\eta{\mu^2\Lambda\over \rho_0}},\ \
\eta={6m_5^6\over \rho_0\mu^2} ~~~(0<\eta\leq 1),
 \en
\be \rho_0=m_\lambda^4+6{m_5^6\over \mu^2}.
 \en
 $\Lambda$ is defined as
 \be
 \label{lam}
\Lambda={1\over 2} ({}^{(5)}\Lambda+{1\over 6}\kappa_5^4\lambda^2),
 \en
 where ${}^{(5)}\Lambda$ is the 5-dimensional cosmological constant
 in the bulk, $\kappa_5$ is the 5-dimensional Newton constant, and
 $\lambda$ is the brane tension.
Note that here there are three mass scales, $\mu$,
$m_\lambda=\lambda^{1/4}$ and $m_5=\kappa_5^{-2/3}$.  ${\cal E}_0$
is a constant related to Weyl radiation. Since we are interested in
the inflation dynamics of the model, as usual, we neglect the
curvature term and dark radiation term in what follows. Also in this
paper we restrict ourselves in the Randall-Sundrum critical case,
that is $\Lambda={1\over 2} ({}^{(5)}\Lambda+{1\over
6}\kappa_5^4\lambda^2)=0$. Then the Friedmann equation (\ref{fried})
can be rewritten as
 \be
 \label{newfried}
 H^2=\frac{1}{3\mu^2}\left[\rho+\rho_0+\epsilon \rho_0 (1
 +\frac{2\eta\rho}{\rho_0})^{1/2}\right].
 \en
 Because only in the branch $\epsilon=-1$ the inflation exits
 spontaneously, we only consider this branch from now on.
 In the ultra high energy limit where
 $ \rho \gg \rho_0 \gg m_{\lambda}^4$, the Friedmann equation
 (\ref{newfried}) is
 \begin{equation}
 \label{high}
 H^2 = \frac{1}{3\mu^2}\left( \rho +\epsilon \sqrt{2\rho
 \rho_0}\right).
 \end{equation}
 This describes a four dimensional gravity on the brane. In the
 intermediate energy region where $\rho \ll \rho_0 $ but $\rho \gg
 m_{\lambda}^4$, for the branch with $\epsilon =-1$, the Friedamnn
 equation changes to
 \begin{equation}
 \label{mid}
 H^2  = \frac{m_{\lambda}^4}{18m_5^6}\left( \rho
+\frac{\rho^2}{2m_{\lambda}^4} -\frac{\mu^2
m_{\lambda}^4}{6m_5^6}\rho -\frac{\mu^2}{4m_5^6}\rho^2\right).
 \end{equation}
 And in low energy limit $\rho \ll m_{\lambda}^4 \ll \rho_0$,
 Friedmann equation (\ref{newfried}) becomes
 \begin{equation}
 \label{low}
 H^2 = \frac{1}{3m^2_p}\left [\rho + {\cal O}\left
(\frac{\rho}{\rho_0}\right )^2\right],
 \end{equation}
 where $m^2_p= \mu^2/(1-\eta)$, $m_p$ is 4-dimensional Planck mass.
 Carefully observe the conditions for (\ref{high}),( \ref{low}),
 and (\ref{mid}), one finds only when
 \be
 \lambda \ll \frac{6m_5^6}{\mu^2},
 \label{5dim}
 \en
 the universe really undergoes a 5-dimensional phase. Otherwise the
 universe will be always  in a 4-dimensional evolution, which is
 without interest. Considering (\ref{5dim}), and only leaving the most
 important term, (\ref{high}), (\ref{mid}),
 and (\ref{low}) are further simplified to
 \be
 \label{high1}
 H^2 = \frac{1}{3\mu^2}\rho,
 \en

 \begin{equation}
 \label{mid1}
 H^2  = \frac{\rho^2}{36m_5^6},
 \end{equation}
 and
 \begin{equation}
 \label{low1}
 H^2 = \frac{\lambda}{18m_5^6}\rho,
 \end{equation}
 respectively.  Clearly, the above 3 equations
 describe effectively 4-dimensional gravity, 5-dimensional gravity,
 and  4-dimensional gravity, respectively. But the corresponding
 gravitational constants are different.
   The dimension transition from ultra high energy region
  to intermediate energy region occurs at
  \bea
  \rho'_{4.5}=\frac{12 m_5^6}{\mu^2},\\
  H'_{4.5}=\frac{2 m_5^3}{\mu^2},
  \ena
  where have used (\ref{high1}) and (\ref{mid1}). Similarly by using
  (\ref{mid1}) and (\ref{low1}),
 the transition from intermediate energy region
  to low energy region occurs at
 \bea
 \label{rho4.5}
  \rho_{4.5}=2 \lambda,\\
  H_{4.5}=\frac{\lambda}{3m_5^3}.
  \label{H4.5}
  \ena
 Consider an inflation scalar field $\phi$ on the brane with exponential potential
 \be
  \label{ipotential}
  V=\widetilde{V}
 e^{-\sqrt{2/p}\frac{\phi}{\mu}},
 \en
 where $\widetilde{V}$ and $p$  are two constants. Introduce
 \be
 u=\frac{V}{\rho_0}.
 \en
 Recall the numerical example given in \cite{cai},
 $\eta =0.99$, $\mu^2 = 0.01 m_p^2 \sim
(10^{17}Gev)^2$, e-folds of the inflation $N=60$ , $p=50$,
 the value of $u$ when the inflation ends $u_{end}=0.05$,
 the value of $u$ when the
 cosmic scale observed today crosses the Hubble horizon
  during inflation $u_i=36$,
  $\frac{\lambda}{\mu^4}=6.7\times
10^{-12}$, $\frac{m_5}{\mu}=0.02$,
$\frac{{}^{(5)}\Lambda}{\mu^2}=6.7\times10^{-14}$.

 Now we check the inflationary process of this example. First,
 \be
 \frac{\lambda \mu^2}{6m_5^6}=0.017 \ll 1,
 \en
 hence the evolution of the universe really undergoes 4-dimensional stage,
 5-dimensional stage, and then the other 4-dimensional stage.
 Second,
 \be
 u_i=36\gg 1,
 \en
 therefore the universe inflates in a 4-dimensional stage when the
 cosmic scale observed today crosses the Hubble horizon
  during inflation. Finally
  \be
  u_{end}=0.05\ll 1,~~~ \frac{\rho}{m_{\lambda}^4}=58.3\gg 1,
  \en
  hence the inflation phase exits in a 5-dimensional stage.

  Then, in the
  model without curvaton field, the
  universe enters a kinetic period, in which the energy density
  decreases very fast, which soon restores the universe to be a 4-dimensional
  again before nucleosynthesis. But, as we have mentioned, such a
  scenario brings several serious problems because the gravitational
  particle production is far from efficient.
 \section{the curvaton reheating}
 In this section we shall explore the permitted parameter region of
 the curvaton field which is responsible for particle production and for the structure
 formation.
  \subsection{The dynamics of the curvaton field}
  Curvaton is a  new mechanism for the primordial curvature
  perturbation generation suggested in literatures in recent years.
   In contrast with the usual inflaton reheating process,
  the inflaton need not roll slowly.  The inflation and reheating
  are charged by different fields, such that many hopeful inflation models survive.
  We here assume that the inflaton has no interactions with inflaton except the gravitational
  coupling. Hence, similar to the inflaton, the equation of motion of curvaton
field $\sigma$
  can be written as
  \be
  \ddot{\sigma}+3H\dot{\sigma}+U'(\sigma)=0.
  \label{motioncur}
  \en
  For simplicity, we assume $U(\sigma)=\frac{1}{2}m^2\sigma^2$. We
  start to give a brief description of the dynamical evolution of
  the curvaton field.
  First, the curvaton coexists with
  the inflaton field throughout the inflationary phase, during which the inflaton
 energy density dominates the energy density of curvaton. Because
  the curvaton is effectively massless before oscillation, $\sigma$ keeps
  at its initial value $\sigma_i$, where the subscript $i$ denotes the value
  when the cosmic scale observed today crosses the Hubble horizon
  during inflation \cite{curvaton}.  The next stage begins when the curvaton
  field becomes to oscillate, and this should happen
  at the kinetic epoch, otherwise its fluctuation can be suppressed
  by the fluctuation of inflaton.

  For the sake of preventing a stage of
 curvaton-driven inflation, the universe must be still dominated by inflaton
 till this time.   This condition imposes a constraint
 on the initial values the curvaton field. In this period
 the curvaton evolves as pressureless dust because its mean kinetic energy
 equals the mean potential energy. From (\ref{motioncur}) we see
 the curvaton becomes to oscillate when $H\backsimeq m$, while the universe
 can be effectively 5-dimensional or 4-dimensional. According
 to our assumption the universe is still dominated by inflaton,
 which behaves as stiff matter in kinetic epoch.
 Therefore we obtain, for the case of oscillating in 5-dimensional
 stage,
 \be
 m/H_{\rm kin}=a_{\rm kin}^6/a_{\rm osc}^6,
 \label{osc5}
 \en
 where kin stands for the value when the universe exits from
 the inflationary phase and  enters the kinetic energy dominated
 epoch. And for the case of oscillating in 4-dimensional
 stage
 \be
 m/H_{\rm kin}=\frac{a_{\rm kin}^6}{a_{\rm 4.5}^6}\frac{a_{4.5}^3}{a_{\rm osc}^3},
 \label{osc4}
 \en
 where $4.5$ stands for the transition point from a 5-dimensional
 stage to a 4-dimensional stage.
 In the oscillating period the energy density of the curvaton field evolves as
 dust,
 \be
 \rho_{\sigma}=\rho_{\sigma \rm i}a_{\rm osc}^3/a^3,
 \label{rhosigma}
 \en
 where $\rho_{\sigma \rm i}=m^2\sigma_{\rm i}^2/2$.
  The third stage is that of curvaton decay
 into radiation, which happens when the decay parameter $\Gamma=H$.
 Here we adopt a sudden decay approximation,
 which is a quite good approximation in curvaton models \cite{sudden}.
    We see there are three free parameters of the
 curvaton field: the initial
 value of the field $\sigma_i$; the mass $m$; and the decay energy scale $\Gamma$.

If the curvaton oscillates in 5-dimensional stage, it may decay in
5-dimensional stage or 4-dimensional stage.
  If the curvaton oscillates in 4-dimensional stage, it must
also decay in
 4-dimensional stage. Hence there are three cases: 1, oscillation in
 5-dimensional stage, decay in 5-dimensional stage; 2,
 oscillation in 5-dimensional stage, decay in 4-dimensional stage;
 and 3, oscillation in 4-dimensional stage, decay in 4-dimensional
 stage. Each of the three cases includes two subcases to be considered,
depending whether the curvaton field decays before or after it
 becomes the dominant component of the universe. In the following
 subsections we shall discuss these cases one by one.
 \subsection{Case 1: oscillation and decay in 5-dimensional stage}

 \begin{figure*}
 \centering
 \includegraphics[totalheight=3in, angle=-90]{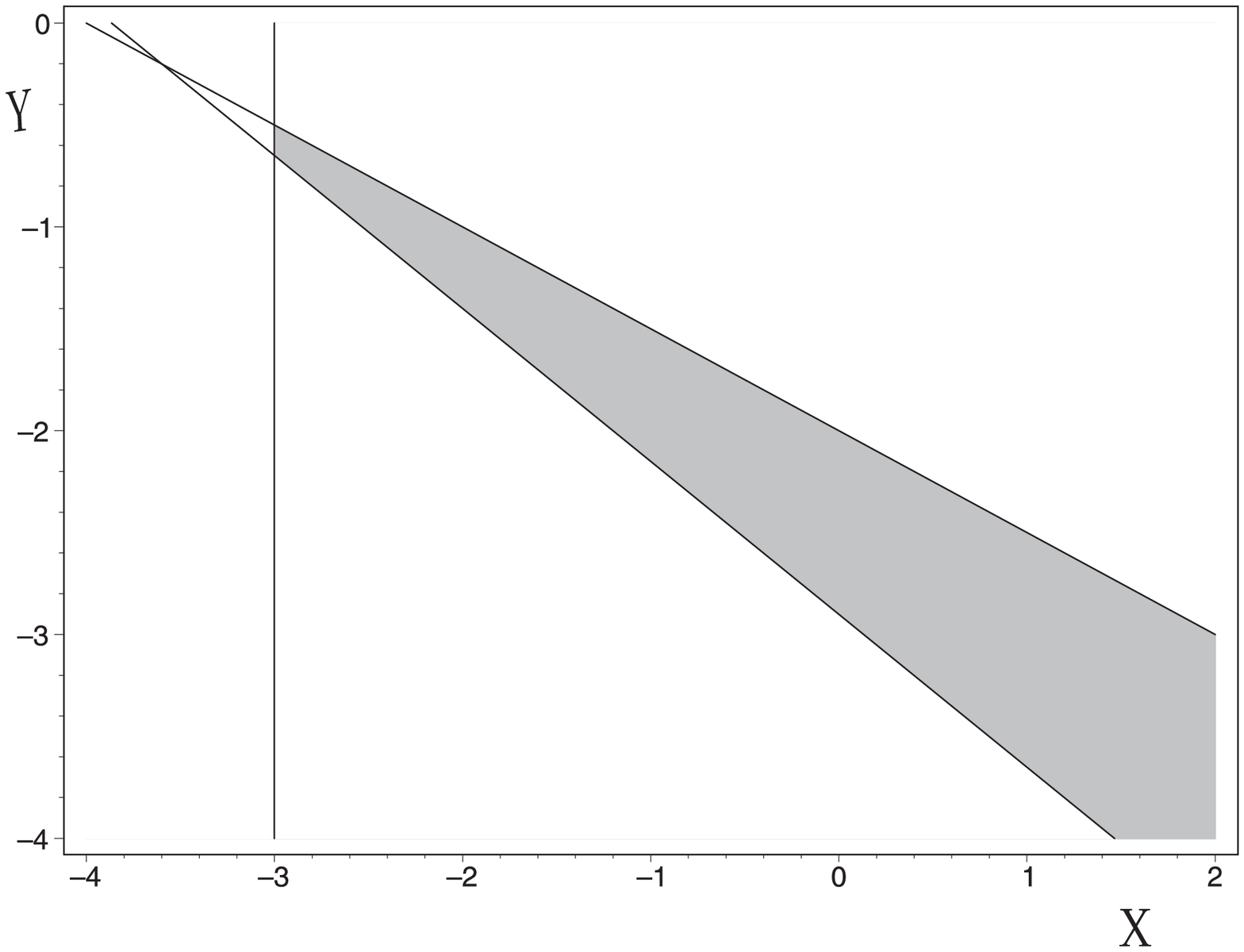}
 \includegraphics[totalheight=3in, angle=-90]{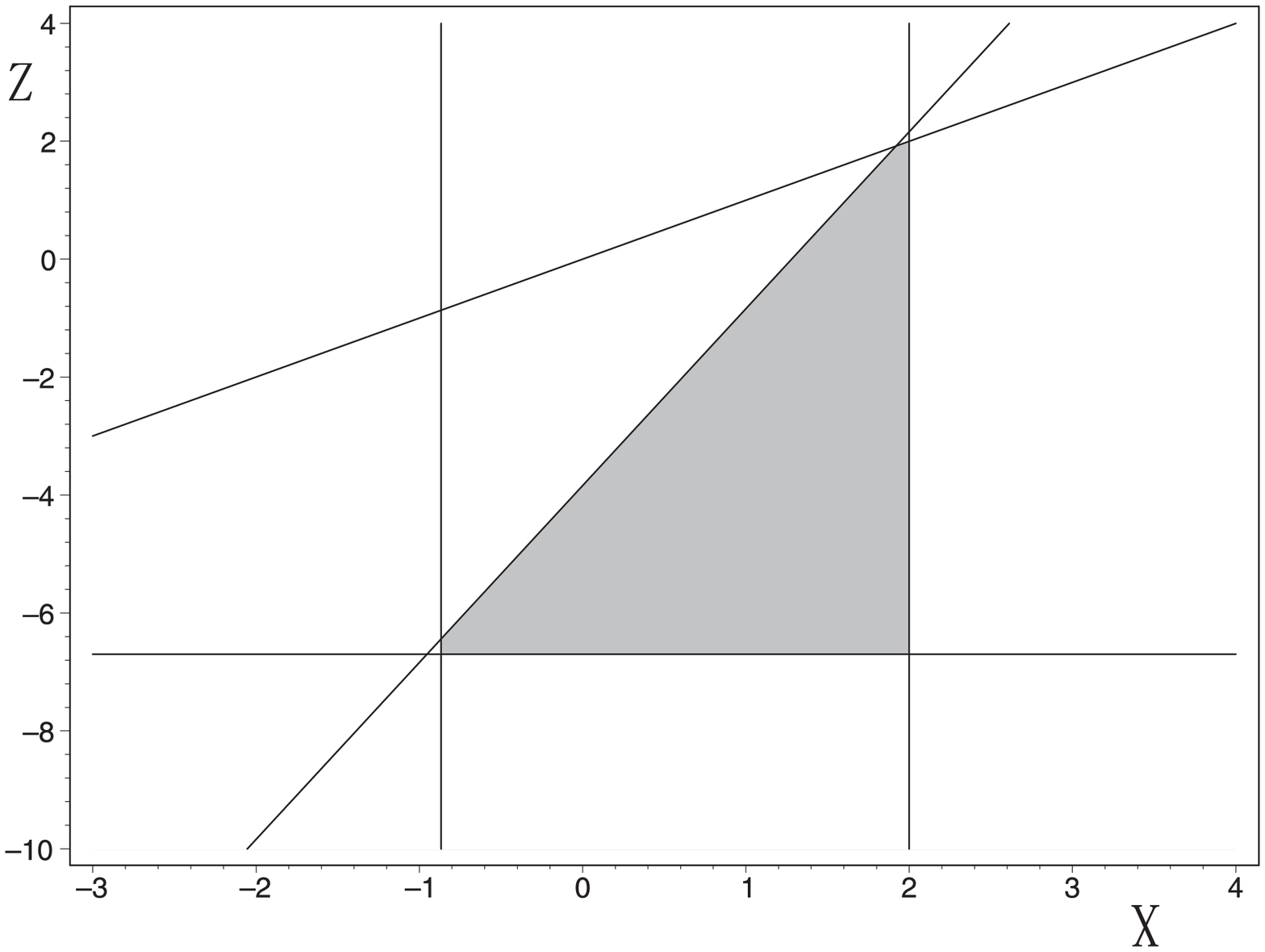}
 \caption{The subcase decay after domination of case 1. In this and all
 the following figures we adopt the numerical example of the parameters in warped DGP
 model given in \cite{cai}, where $\eta =0.99$, $\mu^2 = 0.01 m_p^2 \sim
(10^{17}Gev)^2$,  $\frac{\lambda}{\mu^4}=6.7\times 10^{-12}$,
$\frac{m_5}{\mu}=0.02$,
$\frac{{}^{(5)}\Lambda}{\mu^2}=6.7\times10^{-14}$, and the shadowed
region represents the permitted region. (a) $y$ versus $x$, in which
we take $z=-3$. (b)~$z$ versus $x$, in which we take $y=-3$. }
 \label{}
 \end{figure*}

 \begin{figure*}
 \centering
 \includegraphics[totalheight=3in, angle=-90]{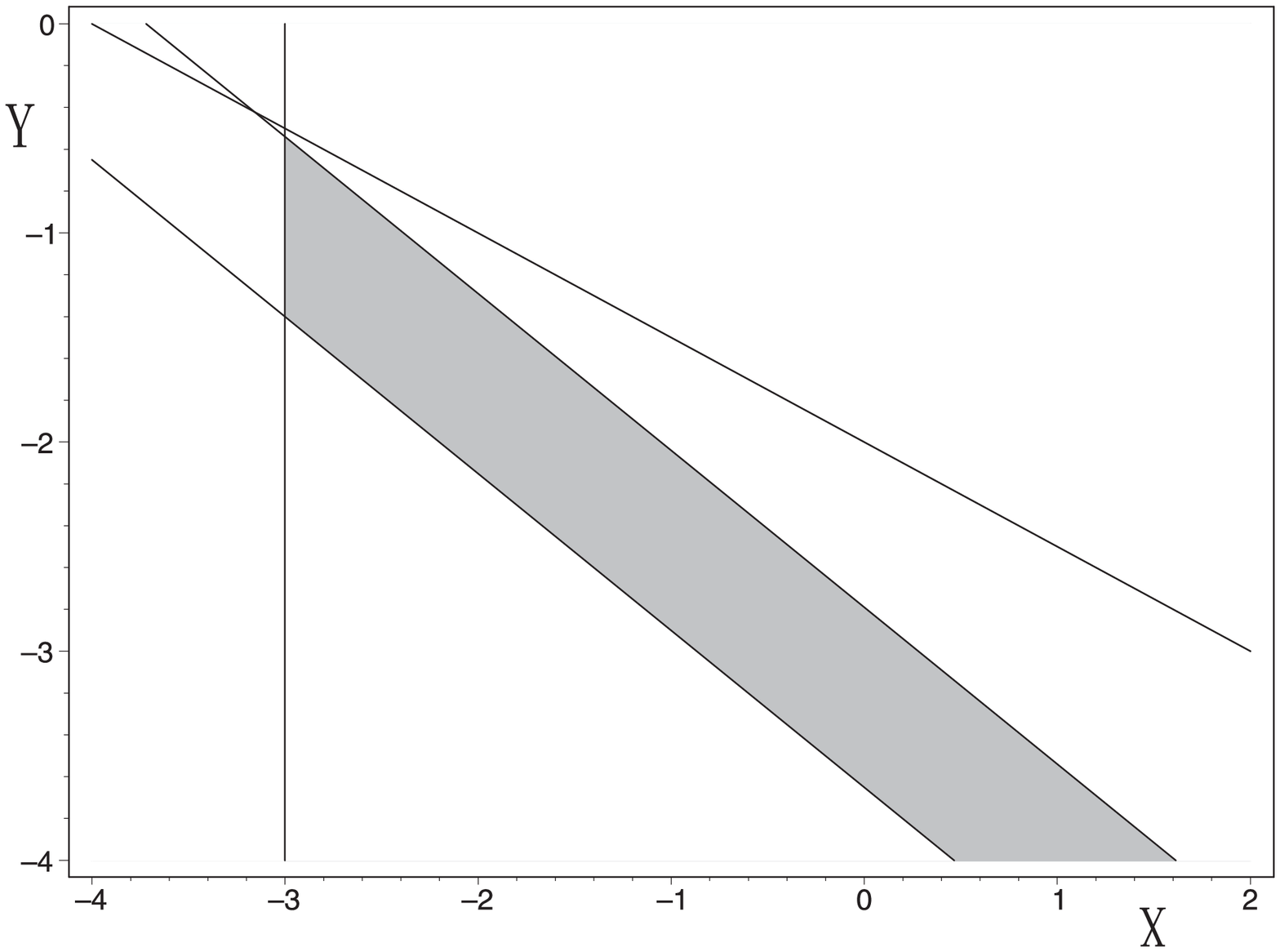}
 \includegraphics[totalheight=3in, angle=-90]{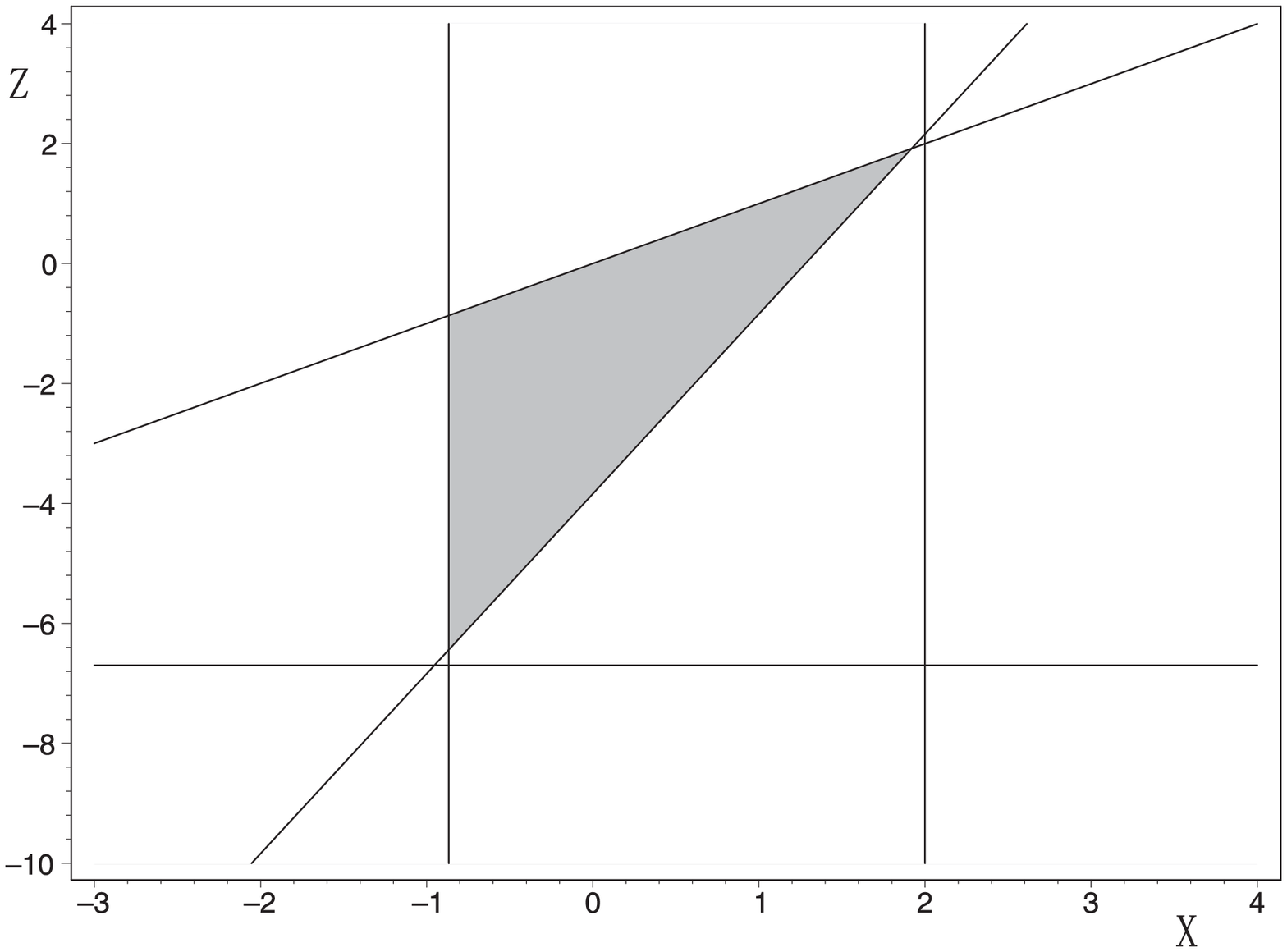}
 \caption{The subcase decay after domination of case 1. (a) $y$ versus $x$, in which
  we take $z=-3$.  (b)~ $z$ versus $x$, in which we take $y=-3$. }
 \label{}
 \end{figure*}

 In this subsection we present the constraints on the parameter space
 of the curvaton in the case of the curvaton oscillates and decays
 in 5-dimensional stage.
 First, we should ensure that the inflaton dominates the evolution
 of the universe when the curvaton starts to oscillate, which
 implies
 \be
 \left. \frac{\rho_{\sigma}}{\rho_{\phi}}\right|_{\rm osc}<1,
 \en
 where
 \be
 \rho_{\sigma}=\rho_{\sigma \rm i}=m^2\sigma_{\rm i}^2/2,
 \en
 since we have assumed the $\sigma$ is effectively a constant
 before oscillation, and by using (\ref{mid1}),
 \be
 \rho_{\phi}=6m_5^3m.
 \en
 So we obtain
 \be
 m\sigma^2_i<12m_5^3.
 \label{control}
 \en
 There are two subcases in this case depending the curvaton
 dominates the universe before or after decay. First, we study the
 curvaton begins to dominate the universe before decay. The events sequence
 is as follow: the curvaton starts to oscillate, the energy density
 of curvaton $\sigma$ equals the energy density of inflaton $\phi$,
 the curvaton decays, the universe transits from 5-dimensional phase
 4-dimensional phase, the nucleosynthesis happens. One can translate this
 sequence into equation
 \be
 m>H_{\rm eq1}>\Gamma>H_{4.5}>H_{\rm nuc},
 \label{c1}
 \en
 where $H_{\rm eq1}$ denotes the value of the Hubble parameter when
  the density of curvaton equals the density of inflaton in 5-dimensional stage,
 $H_{\rm nuc}=10^{-41}m_{\rm Pl}$ denotes the value of the Hubble parameter when
 nucleosynthesis happens.
 The energy density of inflaton evolves as
 \be
 \rho_{\phi}=\rho_{\rm kin}\frac{a_{\rm kin}^6}{a^6}=6m_5^3H_{\rm
 kin} \frac{a_{\rm kin}^6}{a^6}.
 \en
 So we obtain
 \be
  \left. \frac{\rho_{\sigma}}{\rho_{\phi}}\right|_{a=a_{\rm eq1}} =
 \frac{m\sigma_i^2}{2} \frac{a_{\rm osc}^3}{a_{\rm eq}^3}\frac{1}{6m_5^3}
  \frac{a_{\rm eq1}^6}{a_{\rm kin}^6}=1,
 \en
 which yields
 \be
 H_{\rm eq1}=\frac{\sigma_i^4}{4}\frac{m^3}{36m_5^6},
 \en
 where we have used
 \be
 \label{osckin}
 \frac{m}{H_{\rm kin}}\frac{a_{\rm osc}^6}{a_{\rm kin}^6}=1,
 \en
 and
 \be
 H_{\rm eq1}=H_{\rm kin}\frac{a_{\rm kin}^6}{a_{\rm eq1}^6}.
 \en
 Here eq1 labels the time when the density of curvaton equals
 the density of inflaton in 5-dimensional stage. And then (\ref{c1})
 becomes
 \be
 \label{c1'}
 m>\frac{\sigma_i^4}{4}\frac{m^3}{36m_5^6}>\Gamma>\frac{\lambda}{3m_5^3}>H_{\rm
 nuc}.
 \en
 Because of  transitivity (\ref{c1'})  is a fairly strong constraint
 on $m,~\Gamma,~{\rm and}~\sigma_i$, which includes,
 \bea
 \label{1st}
  m &>& \frac{\sigma_i^4}{4}\frac{m^3}{36m_5^6},\\
 m &>& \Gamma,\\
 m &>& \frac{\lambda}{3m_5^3},\\
 m &>& H_{\rm nuc},\\
 \frac{\sigma_i^4}{4}\frac{m^3}{36m_5^6}&>&\Gamma, \\
 \frac{\sigma_i^4}{4}\frac{m^3}{36m_5^6}&>&
 \frac{\lambda}{3m_5^3},\\
 \frac{\sigma_i^4}{4}\frac{m^3}{36m_5^6} &>& H_{\rm
 nuc},\\
 \Gamma&>&\frac{\lambda}{3m_5^3},\\
 \Gamma&>&H_{\rm nuc},\\
 \frac{\lambda}{3m_5^3}&>&H_{\rm
 nuc}.
 \label{last}
 \ena
 Note that (\ref{control}) is just (\ref{1st}). They represent the
 same relation between $\rho_{\phi}$ and $\rho_{\sigma}$ to be
 compared at different times.

 We plot a figure to display the parameter space permitted clearly, in
 which we will use the numerical example in \cite{cai}. In this
 example
 \be
 \frac{\lambda}{3m_5^3 H_{\rm nuc}}=3\times 10^{34}>1,
 \en
 hence equations (\ref{1st})-(\ref{last}) reduce to
 \bea
 \label{11st}
  m &>& \frac{\sigma_i^4}{4}\frac{m^3}{36m_5^6},\\
 m &>& \Gamma,\\
 m &>& \frac{\lambda}{3m_5^3},\\
  \frac{\sigma_i^4}{4}\frac{m^3}{36m_5^6}&>&\Gamma, \\
 \frac{\sigma_i^4}{4}\frac{m^3}{36m_5^6}&>&
 \frac{\lambda}{3m_5^3},\\
 \Gamma&>&\frac{\lambda}{3m_5^3}.
 \label{last1}
  \ena
  Define
  \bea
  x=\log\frac{m}{\mu},\\
  y=\log\frac{\sigma_i}{\mu},\\
  z=\log\frac{\Gamma}{\mu}.
  \ena
 Substitute the values of the parameters in warped DGP model in this
 example, (\ref{11st})-(\ref{last1}) become
 \bea
 x+2y&<&-4,\\
 x&>&-6.55,\\
 x&>&z,\\
 4y+3x&>&z-8.16,\\
 4y+3x&>&-14.6,\\
 z&>&-6.55,
 \ena
 respectively. We show the permitted parameter region in fig. 1.

 Now we turn to subcase 2: curvaton dominates the universe after decay.
 The event sequence is slightly different from the above subcase: the curvaton starts to oscillate,
 the curvaton decays, the energy density
 of curvaton $\sigma$ equals the energy density of inflaton $\phi$,
 the universe transits from 5-dimensional phase
 4-dimensional phase, the nucleosynthesis happens, or by equation
 \be
 m>\Gamma>H_{\rm eq1}>H_{4.5}>H_{\rm nuc}.
 \label{c2}
 \en
 Just mimicking the discussion of the last subcase, we plot the
 permitted parameter region in fig. 2.

 \subsection{Case 2: oscillation in 5-dimensional stage, decay in 4-dimensional stage}

 \begin{figure*}
 \centering
 \includegraphics[totalheight=3in, angle=-90]{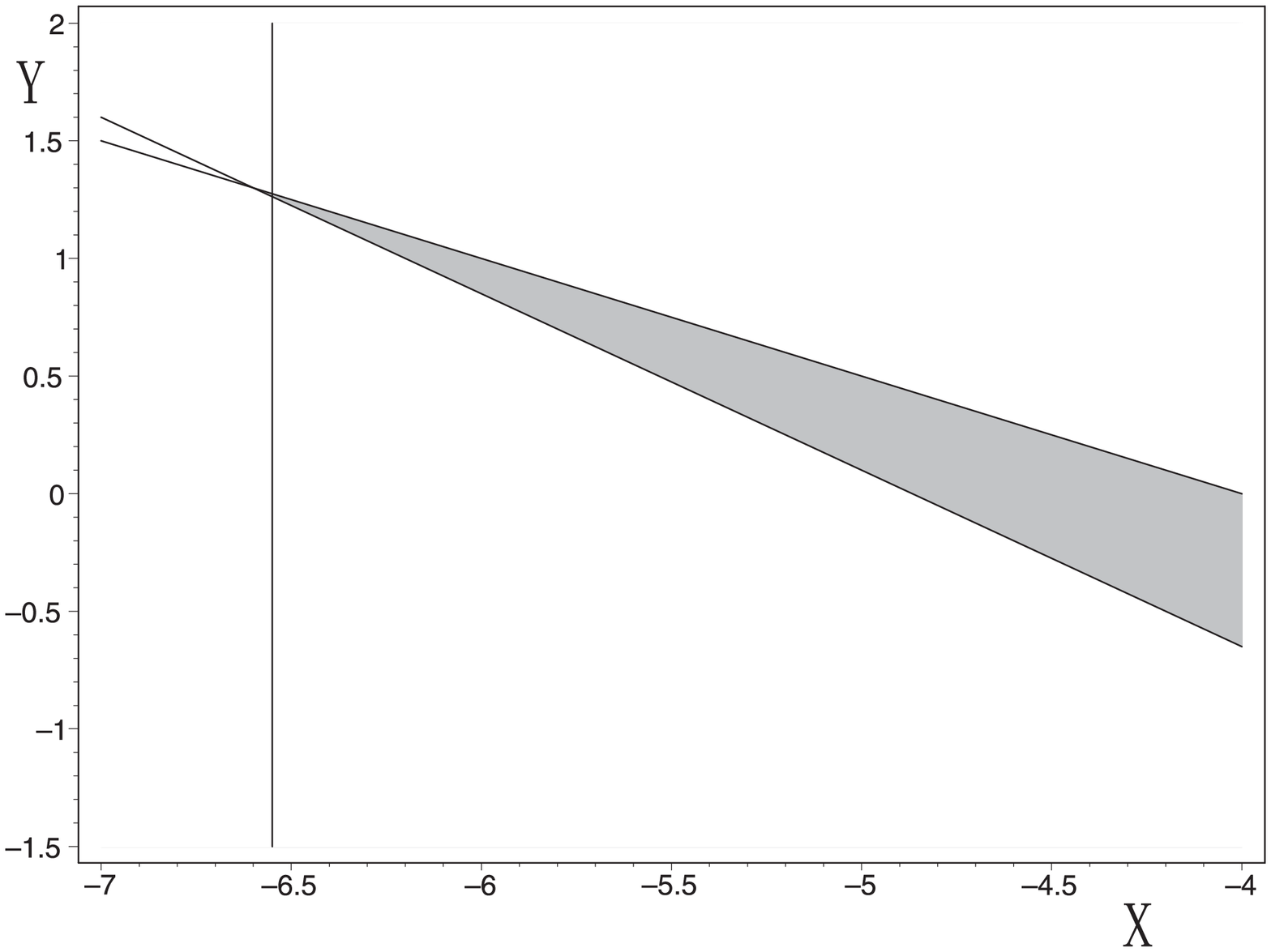}
 \includegraphics[totalheight=3in, angle=-90]{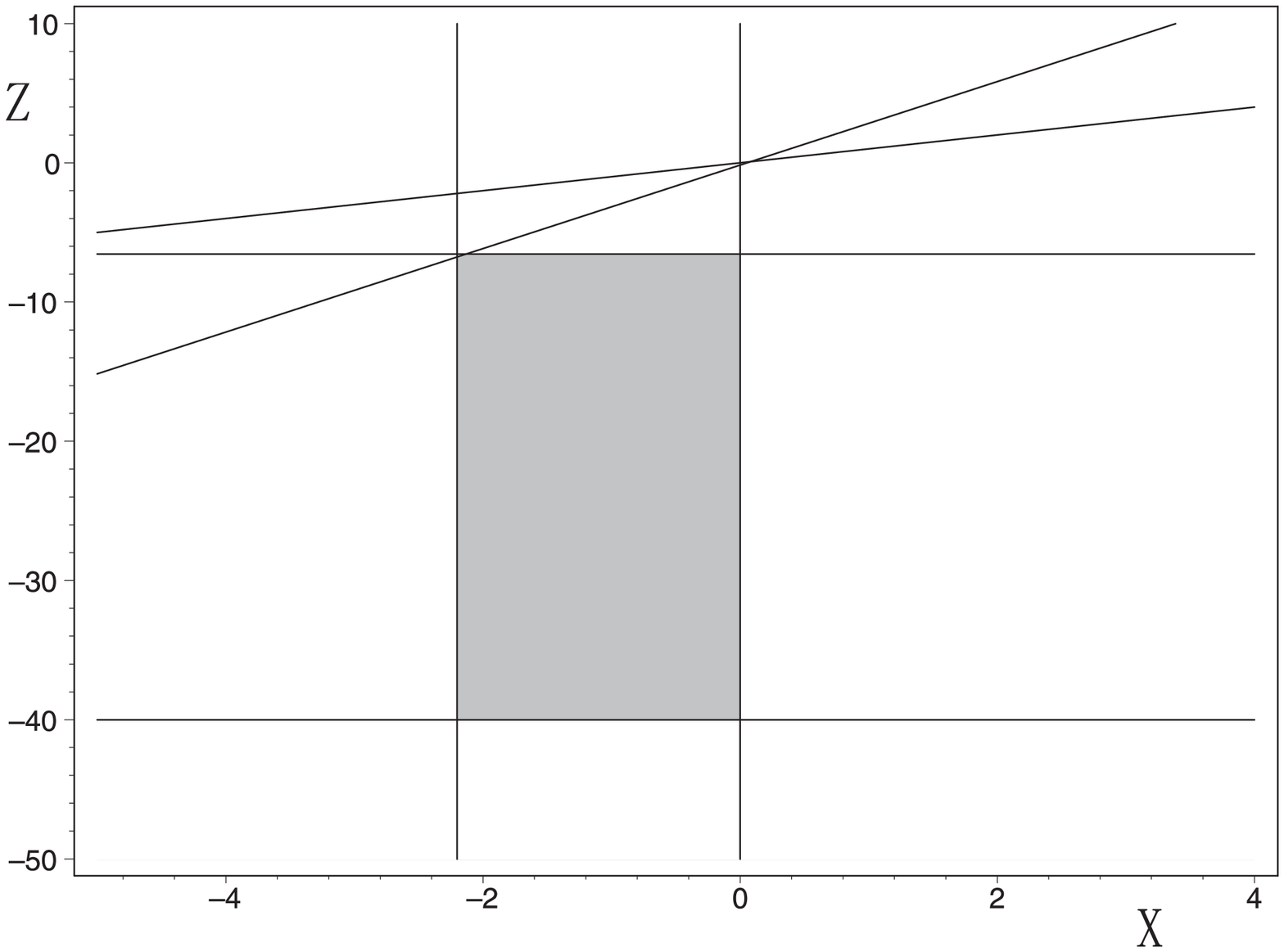}
 \caption{The subcase decay after domination of case 2, where the energy density of curvaton
 equals the energy density of inflaton happening in 5-dimensional stage. (a) $y$ versus $x$, in
 which we set $z=-10$. (b)~$z$ versus $x$, in which we set $y=-2$. }
 \label{}
 \end{figure*}

 \begin{figure*}
 \centering
 \includegraphics[totalheight=3in, angle=-90]{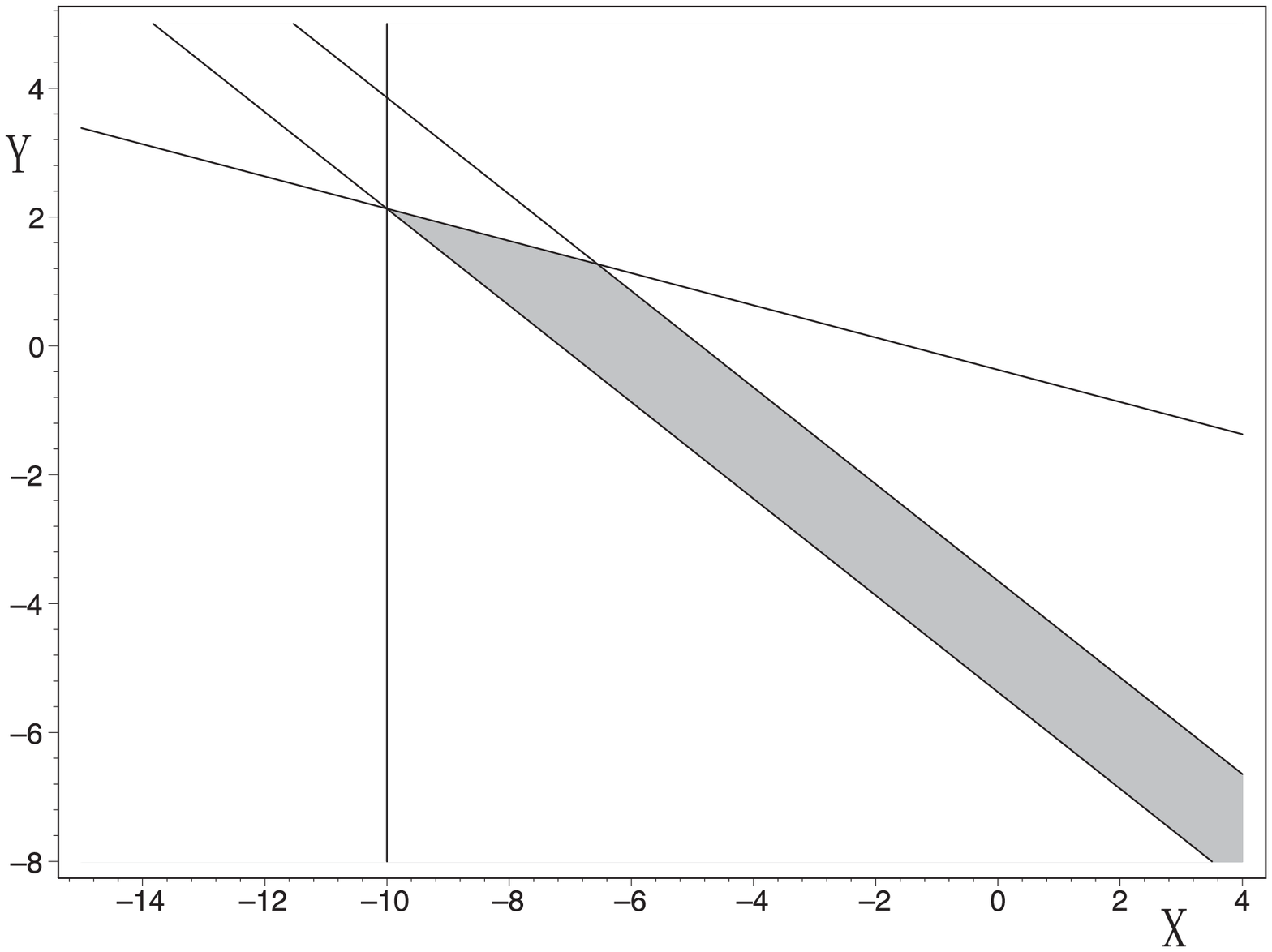}
 \includegraphics[totalheight=3in, angle=-90]{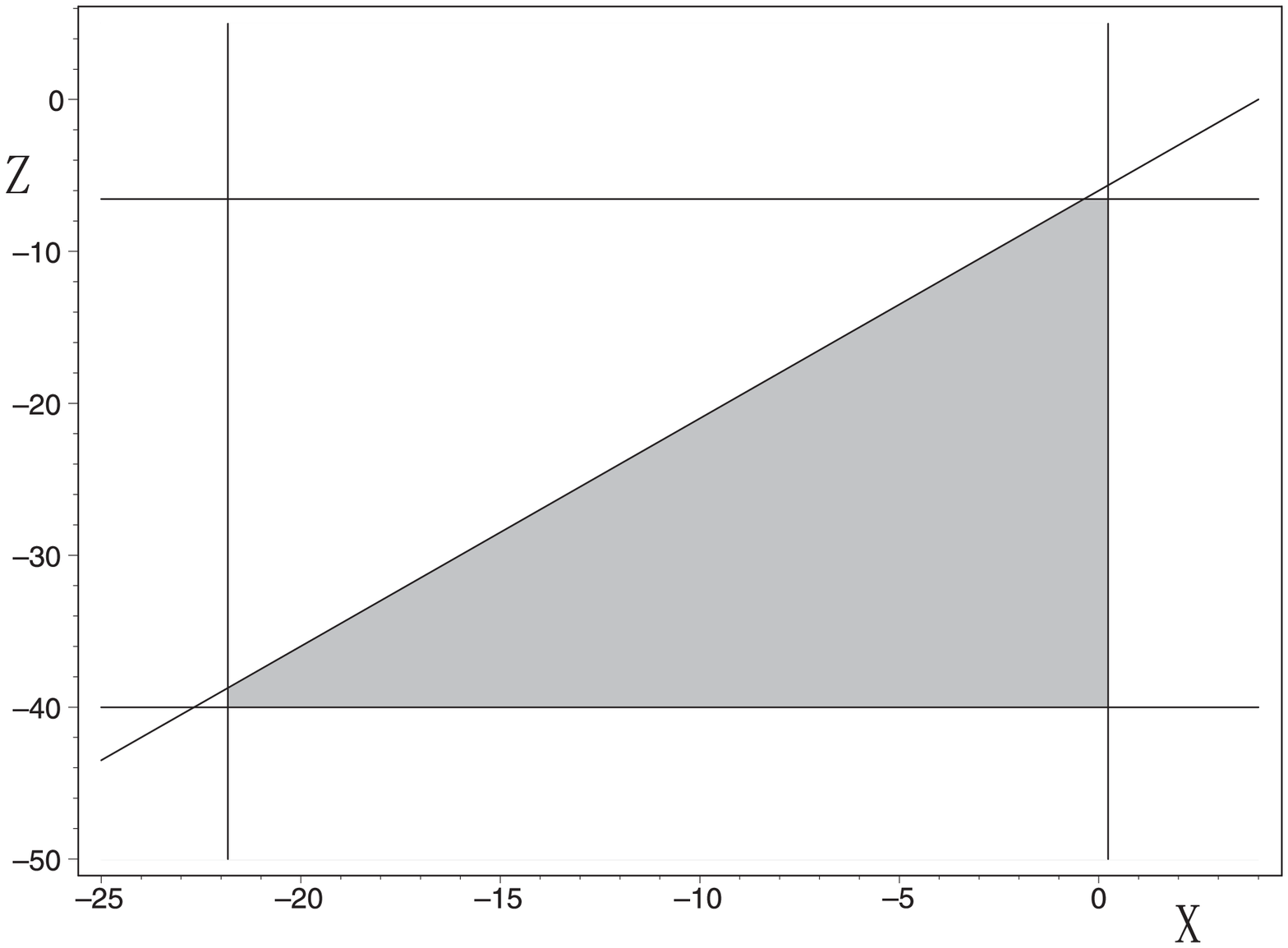}
 \caption{ The subcase decay after domination of case 2, where the energy density of curvaton
 equals the energy density of inflaton happening in 4-dimensional stage. (a) $y$ versus $x$, in which
 we set $z=-10$. (b)~ $z$ versus $x$, in which we set $y=-4$. }
 \label{}
 \end{figure*}

 \begin{figure*}
 \centering
 \includegraphics[totalheight=3in, angle=-90]{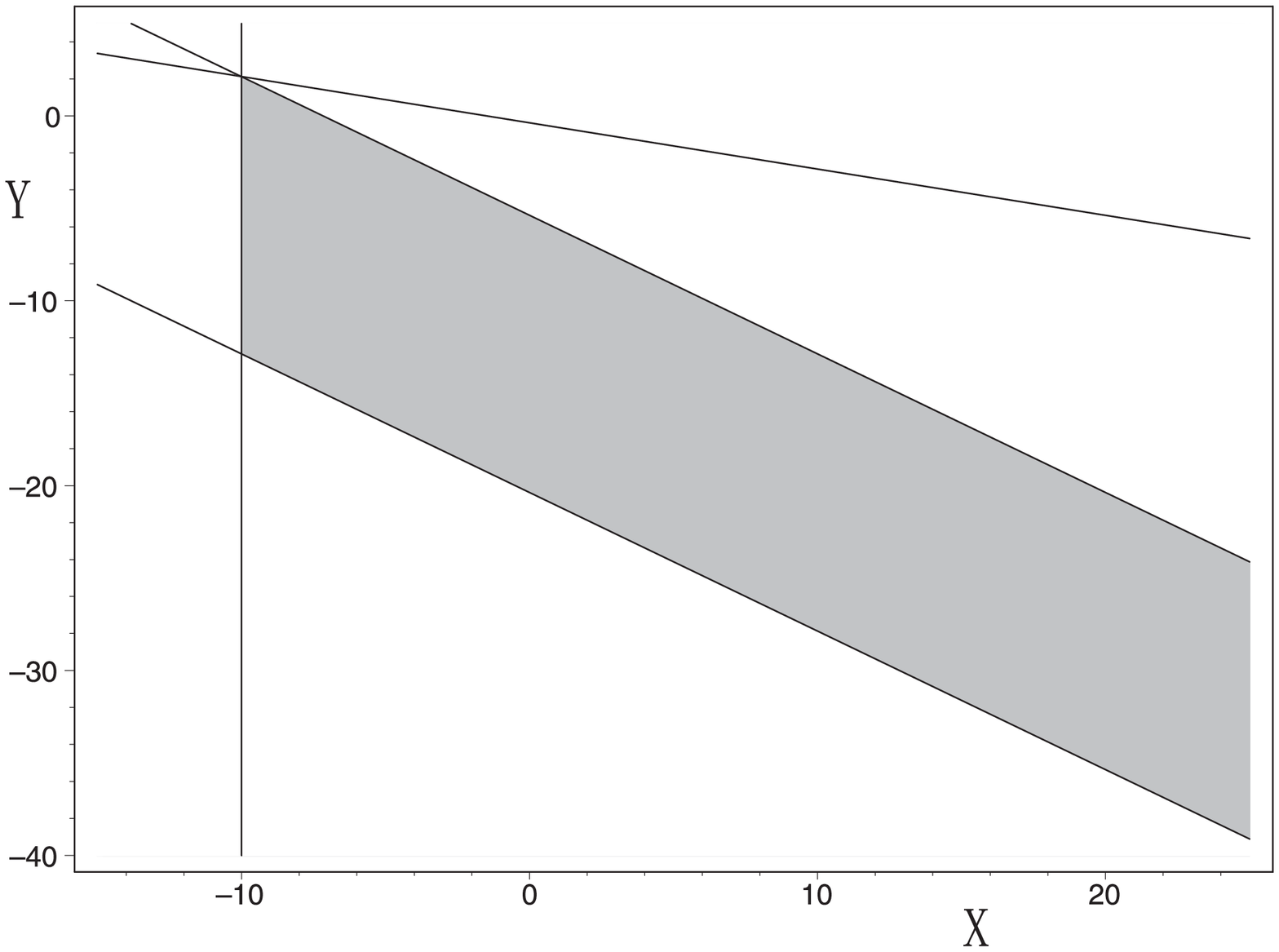}
 \includegraphics[totalheight=3in, angle=-90]{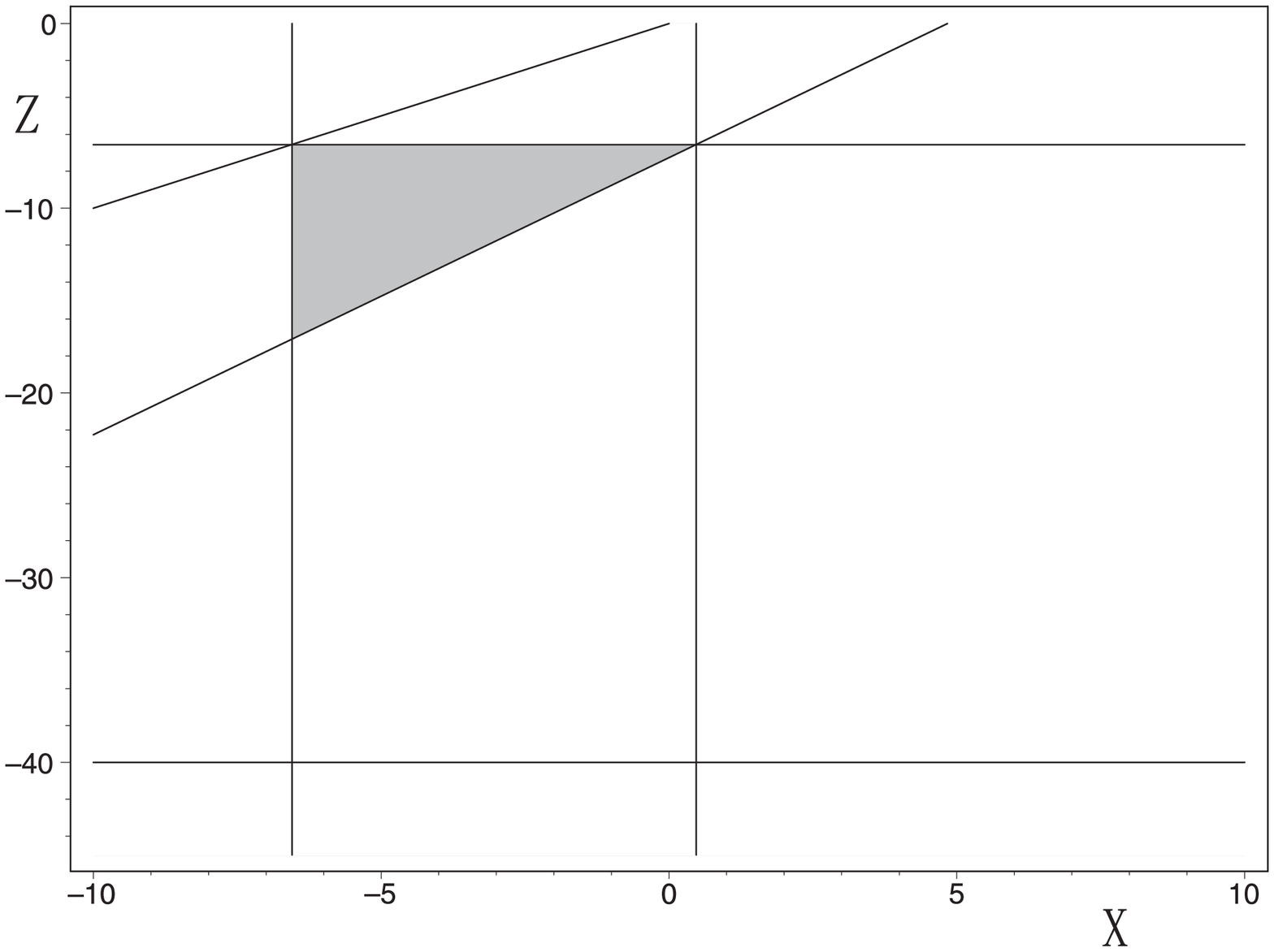}
 \caption{The subcase decay after domination of case 2. (a)  $y$ versus $x$, in which
 we set $z=-10$. (b) $z$ versus $x$, in which we set $y=-4$. }
 \label{}
 \end{figure*}

  In this subsection we deduce the constraints on the parameter space
 of the curvaton in the case of the curvaton oscillates in 5-dimensional stage,
 but decays in 4-dimensional stage.
 The condition that the inflaton dominates the evolution
 of the universe when the curvaton starts to oscillate is the same as in last
 case, which is just (\ref{control}).

 This case also include two subcases  depending the curvaton
 dominates the universe before or after decay. First, we study the
 curvaton begins to dominate the universe before decay. The events sequence
 is as follow: the curvaton starts to oscillate,
 the universe transits from 5-dimensional phase
 4-dimensional phase,  the energy density
 of curvaton $\sigma$ equals the energy density of inflaton $\phi$,
 the curvaton decays, the nucleosynthesis happens; or  the curvaton starts to oscillate,
 the energy density
 of curvaton $\sigma$ equals the energy density of inflaton $\phi$,
 the universe transits from 5-dimensional phase
 4-dimensional phase,
 the curvaton decays, the nucleosynthesis happens. One can translate
 the
 sequences into equation
 \be
 m>H_{\rm eq1}>H_{4.5}>\Gamma>H_{\rm nuc},
 \label{c3}
 \en
 or
 \be
 m>H_{4.5} >H_{\rm eq2}>\Gamma>H_{\rm nuc},
 \label{c4}
 \en
 where eq2 labels the the value of a variable when
  the density of curvaton equals the density of inflaton in latter 4-dimensional
  stage, while the curvaton begins to oscillate in 5-dimensional stage.
    We demonstrate the numerical result of (\ref{c3}) in fig.
  3, with the same parameters of the warped DGP model in the above
  case.

 To investigate (\ref{c4}) we need to find $H_{\rm eq 2}$ in
 advance. Under this situation the curvaton red shifts as
   \be
 \rho_{\sigma}=\rho_{\sigma \rm i}\frac{a_{\rm osc}^3}{a_{4.5}^3}
  \frac{a_{\rm 4.5}^3}{a^3},
  \label{4dimcurvaton}
 \en
 and inflaton red shifts as
  \be
  \label{4diminflaton}
  \rho_{\sigma}=6m_5^3H_{\rm kin}\frac{a_{\rm
  kin}^6}{a_{4.5}^6}\frac{a_{4.5}^6}{a^6}.
  \en
  The Hubble parameter evolves as
  \be
  H=H_{\rm kin}\frac{a_{\rm kin}^6}{a_{4.5}^6}
  \frac{a_{4.5}^6}{a^6}.
  \label{4dimhubble}
  \en
  The energy density of curvaton equals the energy density of inflation
  when $a=a_{\rm eq2}$, which means,
  \be
  \left. \frac{\rho_{\sigma}}{\rho_{\phi}}\right|_{a=a_{\rm eq2}} =1,
 \en
  hereby we derive
  \be
  H_{\rm eq2}=\frac{m\sigma_i^2}{12m_5^3}\sqrt{\frac{\lambda
  m}{3m_5^3}}~,
  \en
  where we have used (\ref{osckin}), which is still valid because the
  curvaton oscillate
  in 5-dimensional stage. From now on we follow the same program in
  case 1.
  (\ref{c4}) converts to ten equations,
  \bea
  \label{1st3}
  m&>&\frac{\lambda}{3m_5^3},\\
  m&>&\frac{m\sigma_i^2}{12m_5^3}\sqrt{\frac{\lambda
  m}{3m_5^3}}~,\\
  m&>&\Gamma,\\
  m&>&H_{\rm nuc},\\
  \frac{\lambda}{3m_5^3}&>&\frac{m\sigma_i^2}{12m_5^3}\sqrt{\frac{\lambda
  m}{3m_5^3}}~,\\
  \frac{\lambda}{3m_5^3}&>&H_{\rm nuc},\\
  \frac{\lambda}{3m_5^3}&>&\Gamma,\\
 \frac{m\sigma_i^2}{12m_5^3}\sqrt{\frac{\lambda
  m}{3m_5^3}}&>&\Gamma,\\
 \frac{m\sigma_i^2}{12m_5^3}\sqrt{\frac{\lambda
  m}{3m_5^3}}&>&H_{\rm nuc},\\
 \Gamma&>&H_{\rm nuc}.
 \label{last3}
 \ena
 We show the permitted parameter region of the curvaton field in
 this subcase in fig. 4.

  The other subcase is that the curvaton starts to
  dominate the universe after decay, which means,
   \be
 m>H_{4.5} >\Gamma>H_{\rm eq2}>H_{\rm nuc}.
  \label{c5}
  \en
 Completely following the same procedures of the above subcase,
 we draw fig. 5 to show the permitted region of this subcase.

 \subsection{Case 3: oscillation and decay in 4-dimensional stage}

 \begin{figure*}
 \centering
 \includegraphics[totalheight=3in, angle=-90]{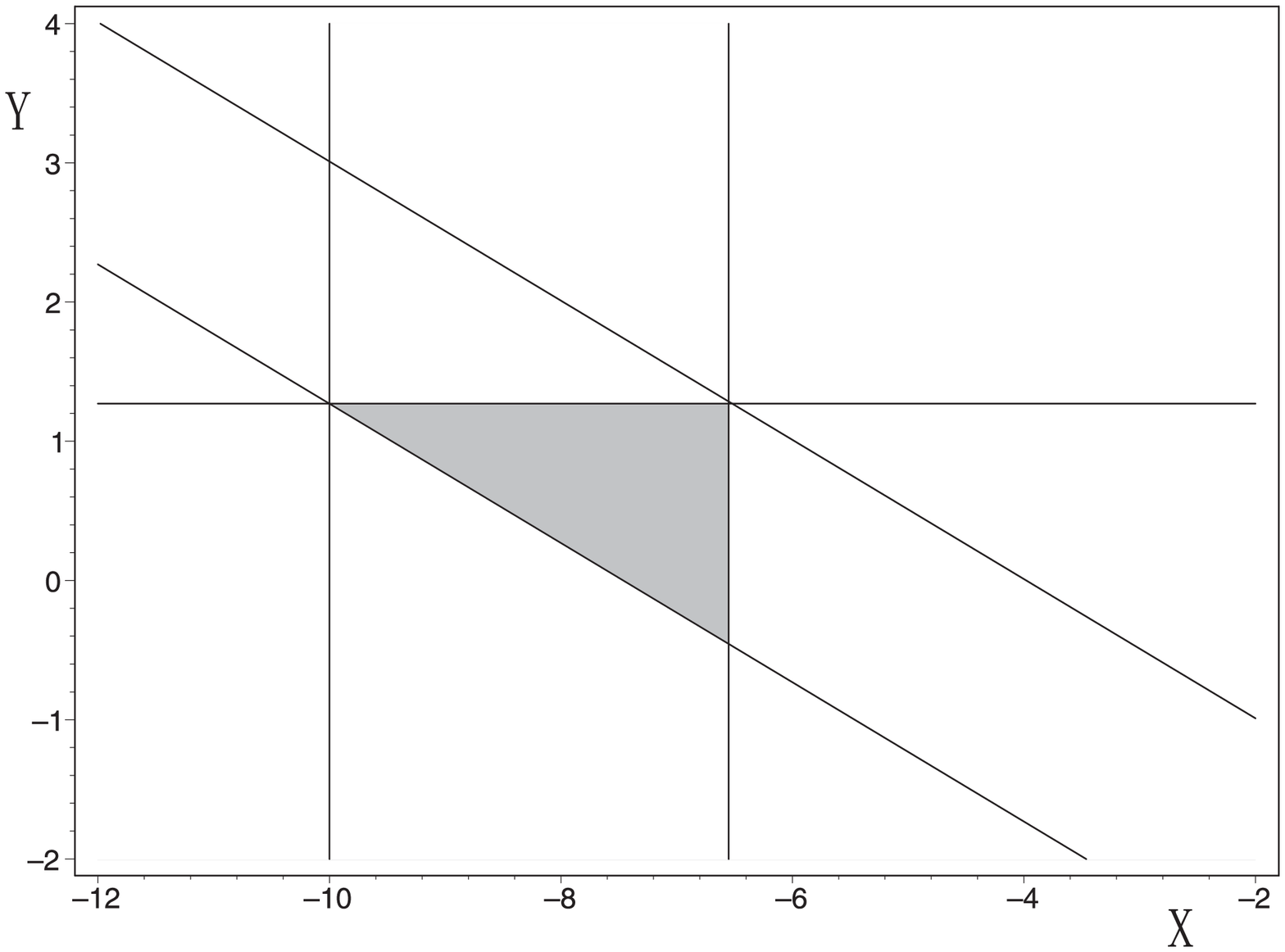}
 \includegraphics[totalheight=3in, angle=-90]{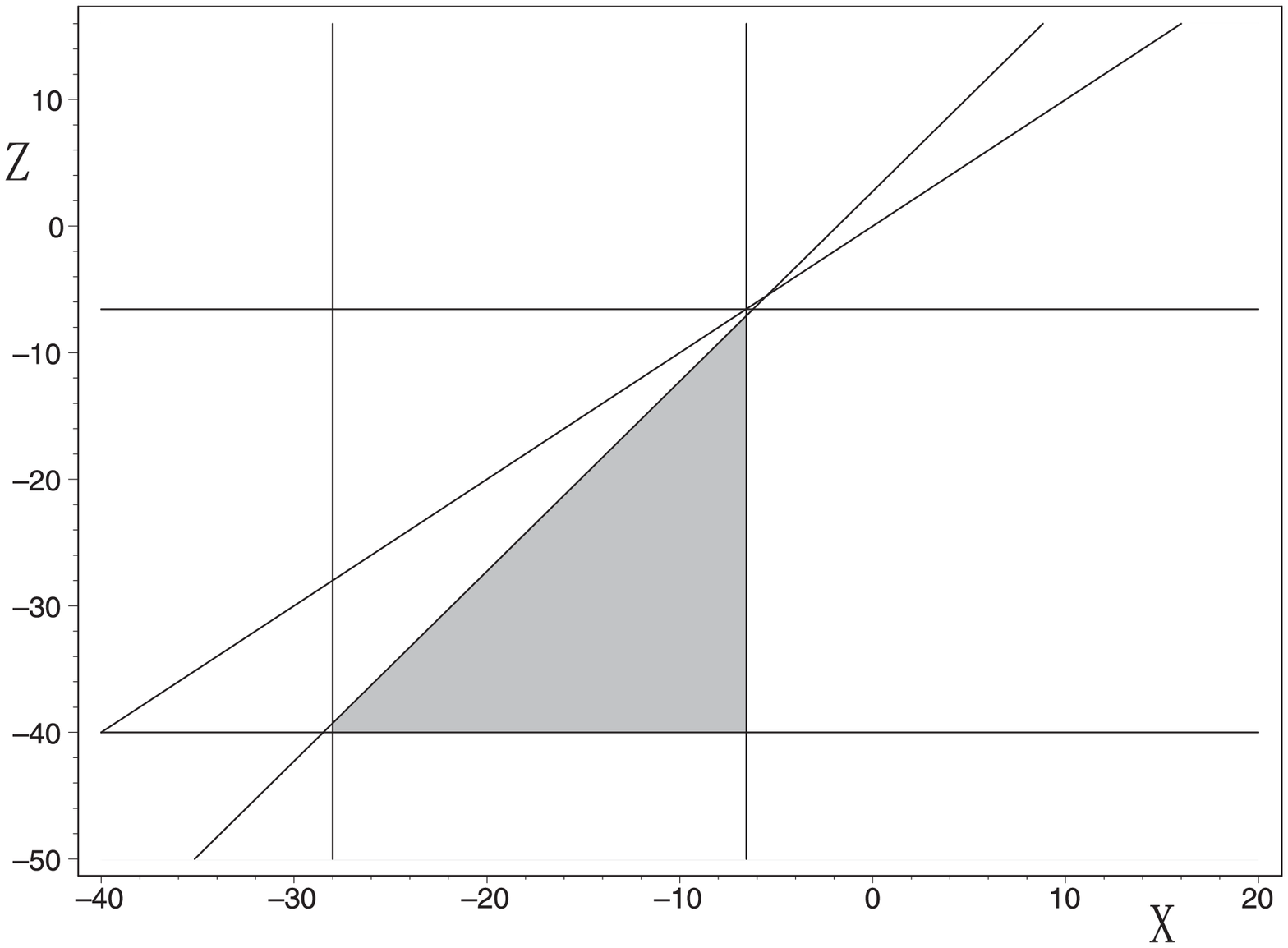}
 \caption{The subcase decay after domination of case 3. (a) $y$ versus $x$, where we fix $z=-10$.
  (b) $z$ versus $x$, where we fix $y=1$. }
 \label{}
 \end{figure*}

 \begin{figure*}
 \centering
 \includegraphics[totalheight=3in, angle=-90]{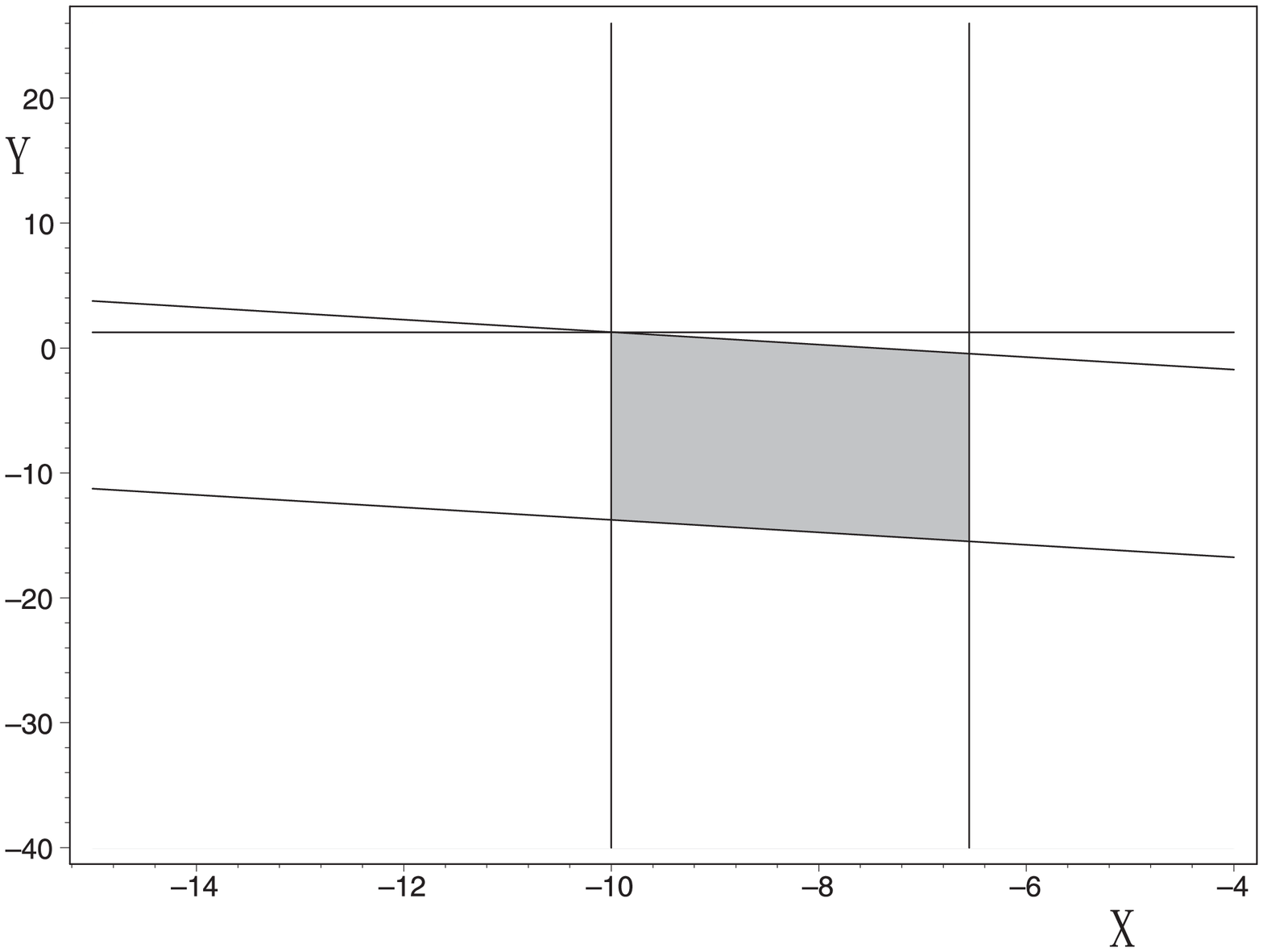}
 \includegraphics[totalheight=3in, angle=-90]{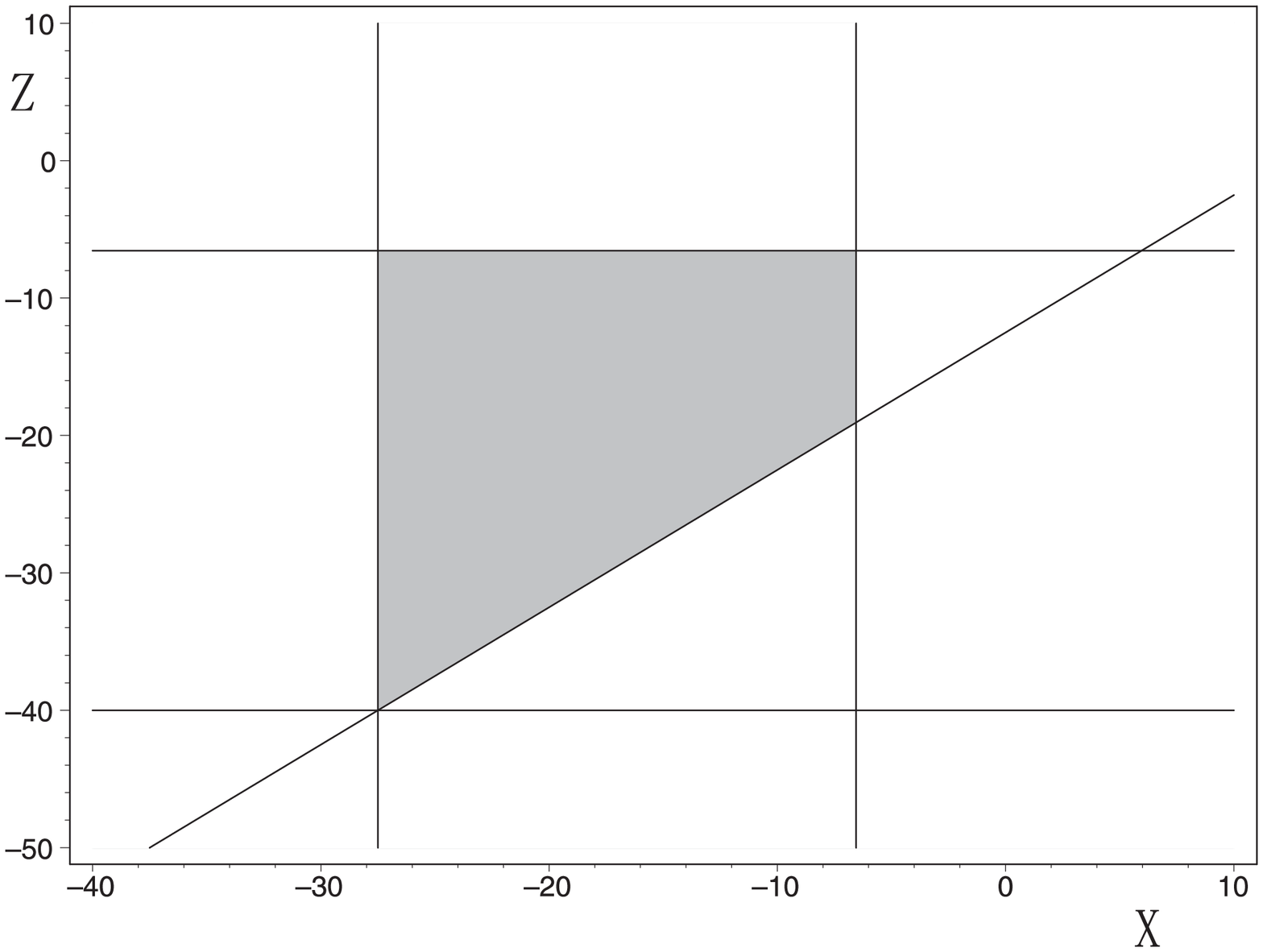}
 \caption{The subcase decay before domination of case 3. (a) $y$ versus $x$, where we fix $z=-10$.
  (b) $z$ versus $x$, where we fix $y=-5$. }
 \label{}
 \end{figure*}

 Surely, the curvaton can oscillate and decay after the universe
 arriving at the 4-dimensional phase. First we study the condition
 that the universe is dominated by inflaton when the curvaton
 starts to oscillate. In this latter 4-dimensional stage,
 the energy density of inflation red shifts according to
 (\ref{low1}),
 \be
 \rho_{\phi}=\frac{18 m_5^6}{\lambda} H^2.
 \en
 Hence
 \be
 \left. \frac{\rho_{\sigma}}{\rho_{\phi}}\right|_{H=m}<1,
 \en
 from which we derive
 \be
 \sigma_i^2<\frac{36m_5}{\lambda},
 \label{control2}
 \en
 where we have used $\rho_{\sigma}=m^2\sigma_i^2/2$. Therefore we
 can directly get the constraint on the initial value of the
 curvaton if it oscillates and decays in a 4-dimensional stage.

   Also, there are two subcases: one is
 that the curvaton dominates the universe before decay; the other is
 that the curvaton dominates the universe after decay. The events sequence
 of the former subcase: the curvaton starts to oscillate,
  the universe transits from 5-dimensional phase
 4-dimensional phase, the energy density
 of curvaton $\sigma$ equals the energy density of inflaton $\phi$,
 the curvaton decays, the nucleosynthesis happens. The sequence of
 Hubble parameters is
 \be
 H_{4.5}>m>H_{\rm eq3}>\Gamma>H_{\rm nuc},
 \label{c6}
 \en
 where eq3 stands for the the value of a variable when
  the density of curvaton equals the density of inflaton in latter 4-dimensional
  stage, at the same time the curvaton also begins to oscillate in 4-dimensional stage.
  The energy density of the curvaton field decreases as
  (\ref{rhosigma}). And the energy density of the inflaton field
  decreases as
   \be
 \rho_{\phi}=\rho_{4.5}\frac{a_{4.5}^6}{a^6}.
 \en
 At the equality point,
 \be
 \left. \frac{\rho_{\sigma}}{\rho_{\phi}}\right|_{a=a_{\rm eq3}} =
 \frac{m^2 \sigma_i^2}{2\rho_{4.5}} \frac{a_{\rm osc}^3}{a_{4.5}^3}
 \frac{a_{\rm eq3}^3}{a_{4.5}^3}=1.
 \label{eq3}
 \en
 Substitute
 \bea
  \frac{a_{\rm osc}^3}{a_{4.5}^3}=\frac{H_{4.5}}{m},\\
  \frac{a_{\rm eq3}^3}{a_{4.5}^3}=\frac{H_{4.5}}{H_{\rm eq3}},
  \ena
 in (\ref{eq3}), we obtain
 \bea
 H_{\rm eq3}=\frac{m \sigma_i^2}{2 \rho_{4.5}}
 H_{4.5}^2=\frac{m\sigma_i^2}{36} \frac{\lambda}{m_5^6},
 \ena
 where we have used (\ref{rho4.5}) and (\ref{H4.5}). So (\ref{c6})
 yields,
 \bea
 \frac{\lambda}{3m_5^3}&>&m,\\
 \frac{\lambda}{3m_5^3}&>&\frac{m\sigma_i^2}{36}
 \frac{\lambda}{m_5^6},\\
 \frac{\lambda}{3m_5^3}&>&\Gamma,\\
 \frac{\lambda}{3m_5^3}&>&H_{\rm nuc},\\
 m&>&\frac{m\sigma_i^2}{36}
 \frac{\lambda}{m_5^6}, \label{same} \\
 m&>&\Gamma,\\
 m&>&H_{\rm nuc},\\
 \frac{m\sigma_i^2}{36}
 \frac{\lambda}{m_5^6}&>&\Gamma, \\
 \frac{m\sigma_i^2}{36}
 \frac{\lambda}{m_5^6}&>&H_{\rm nuc}, \\
 \Gamma&>&H_{\rm nuc}.
 \ena
 Note that (\ref{control2}) is also just (\ref{same}). They represent the
 same relation between $\rho_{\phi}$ and $\rho_{\sigma}$ to be
 compared at different times.
  With the same discussions used before and by the same group of parameters
   of warped DGP model, we obtain fig. 6 to show the permitted
 parameter region  of the curvaton in this subcase.

 The events sequence
 of the latter subcase: the curvaton starts to oscillate,
  the universe transits from 5-dimensional phase
 4-dimensional phase, the curvaton decays, the energy density
 of curvaton $\sigma$ equals the energy density of inflaton $\phi$,
  the nucleosynthesis happens. The sequence of
 Hubble parameters is
 \be
 H_{4.5}>m>\Gamma>H_{\rm eq3}>H_{\rm nuc}.
 \label{c7}
 \en

  We also show the permitted parameter region of curvaton in this
  subcase in fig. 7.
  From figs. $1-7$ we see that generally speaking if the curvaton
  oscillates in a 5-dimensional stage the permitted region of $m,~\sigma_i$
  is open for the value of $x$ can be arbitrarily large if $y$ is small enough,
  while if the curvaton
  oscillates in a 4-dimensional stage the permitted region of
  $m,~\Gamma$ is close. In contrast with the permitted region of $m, ~\sigma$, the
  permitted region of $m,~\Gamma$ always keeps close. So the permitted parameter
  region of the curvaton is much more ample when it oscillates in
  the 5-dimensional stage than in the 4-dimensional stage, that is,
  a reasonable curvaton model is much easier if the curvaton becomes to
  oscillate in the 5-dimensional stage.

  In all of
  these cases there exists viable curvaton model satisfying the
  requirements such as a enough high reheating energy scale,
  sufficient particle generation mechanism for nucleosynthesis
  (comparing to ``gravitational production mechanism'').


 \section{Conclusion and discussion}

  The inflation model on DGP brane is of very interest and attraction
    because the universe can exit the inflationary phase spontaneously
    without any additional mechanism for an exponential potential, which
    is generally a serious problem for the ordinary inflation model with
    an exponential potential. However, this
  model suffer from the problem that the particles generated by gravitation
  is far from efficiency when nucleosynthesis happens.

  In this paper we investigate the curvaton mechanism in warped DGP model. We find it
  can hurdle the inefficient particle production problem with fairly
  ample parameter regions. Because the curvaton may oscillates and
  decay in a 5-dimensional stage or 4-dimensional stage, we discuss
  the 3 cases, say, oscillation in
 5-dimensional stage, decay in 5-dimensional stage;
 oscillation in 5-dimensional stage, decay in 4-dimensional stage;
  and  oscillation in 4-dimensional stage, decay in 4-dimensional
 stage, respectively. We plot figures for every case to show
 permitted parameter regions clearly.

 Other constraints on the parameter of the curvaton field, such as the
 fluctuations for structure formation generated by the curvaton and the primordial
 gravitational wave etc., should be should be further studied in the future work.

 {\bf Acknowledgments.}
 This work was supported by
  the National Natural Science Foundation of China
    , under Grant No. 10533010, by SRF for ROCS, SEM of China,
    and by Program for New Century Excellent Talents in University (NCET).

\end{document}